\documentclass[showpacs,a4paper,prc,aps]{revtex4}

\usepackage[dvips]{graphicx}
\usepackage{color}
\usepackage{amssymb}
\usepackage{amsmath}
\usepackage{amsfonts}
\usepackage{indentfirst}
\RequirePackage{hyperref}
\usepackage{geometry}

\allowdisplaybreaks

\begin{document}

\title{Reaction $e^+e^-\to\omega\pi$ and $\rho(1450)$ and $\rho(1700)$ mesons in quark model}
\author{K.~Kittimanapun}
\email{oou\_kritsada@hotmail.com}
\author{K.~Khosonthongkee}
\author{C.~Kobdaj}
\author{P.~Suebka}
\author{Y.~Yan}
\affiliation{School of Physics, Institute of Science, \\
Suranaree University of Technology,\\
Nakhon Ratchasima, Thailand}
\date{\today}

\begin{abstract}
The investigation in the work of the reaction $e^+e^-\to\omega\pi^0$ in the
$^3P_0$ nonrelativistic quark model reveals that
the $e^+e^-\to\omega\pi^0$ process at the energy region from the
$\omega\pi$ threshold
to 2.0 GeV is dominated by the two-step
process in which the primary $\overline qq$ pair first forms $\rho$ and
$\rho'$ mesons and then the vector mesons decay into $\omega$ and $\pi$.
With $\rho(1450)$ and $\rho(1700)$ mainly in $2S$ and $1D$
states respectively, the experimental data for the cross section of the
reaction $e^+e^-\to\omega\pi^0$ are well produced in the $^3P_0$ quark model.
The work supports the argument that $\rho(1450)$ is mainly a $2S$ meson and
$\rho(1700)$ a $1D$ meson.
\end{abstract}


\maketitle

\section{Introduction}
Decay modes $\rho\pi$ and $\omega\pi$ are among the most important
for the processes of $e^+e^-$ annihilation into hadrons at low
energies, giving mainly the $3\pi$ and $4\pi$ final states,
respectively \cite{lite1,lite2}. These reactions might be used to
study the dynamics of light vector mesons, for example, $\rho'$
and $\omega'$, which may be formed as the intermediate states and
decay then into $\omega\pi$ and $\rho\pi$. The information of
mesons except the lightest ones is still rather rare because of
the lack of high-quality experimental data and also effective
theoretical models. In addition, that $\rho'$ may decay into
$\omega$ ($\rho'\to\omega\pi$) and $\omega'$ to $\rho$
($\omega'\to\rho\pi$) \cite{lite3} adds more uncertainties
to the understanding of the properties of the intermediate states.
The analysis in the work \cite{lite4} confirms that the uncertainties
between the $\omega$-like resonance and $\rho$-like resonance
result in calculations with low accuracy.

Recently, experiments have been set up to study the processes of
$e^+e^-$ annihilation to $\pi^0\pi^0\gamma$ at low energies (below
$2$ GeV). The SND experiments in the energy region $0.6-0.97$ GeV
give information of $\rho$ and $\omega$ intermediate state mesons
\cite{lite5,lite6}. The reaction of
$\omega\pi\to\pi^0\pi^0\gamma$ measured in the center mass
energies $0.92-1.38$ GeV at CMD-2 shows the interference of
$\rho(770)$ meson and $\rho(1450)$ meson, which decays into
$\omega\pi^0$ \cite{lite7}. However, the SND experiment with the
energy up to $1.4$ GeV from the threshold \cite{lite8} revealed
that the experimental cross section can be satisfactorily
understood with two excited states $\rho'(1400)$ and
$\rho''(1600)$ in which the contribution of the $\rho''(1600)$
meason dominates. However, this result contradicts with the
theoretical expectation, where $\rho'$ and $\rho''$ are considered
as $2S$ and $1D ~q\bar q$ states respectively and the $\rho'$
gives a larger contribution.

The intermediate vector mesons in $e^+e^-$ annihilation reactions
at low energies could be simple $\overline qq$ states, mixtures of
$\rho$-like and $\omega$-like mesons, or even hybrid states
($\overline qq$ plus one or more gluons). The idea of exotic meson
(vector hybrid) \cite{lite11} has been proposed, but the
theoretical results are not in line with the the experimental
data. On the other hand, $q\bar q$ structured mesons with
different radial and orbital excitations have been extensively
studied. An earlier work in quark model \cite{lite12} predicted a
series of excited vector mesons, with $\rho(1450)$ and
$\omega(1460)$ being the lowest $\rho$-type state with the
$2\,^3S_1$ excitation which has a large probability to decay into
$\omega\pi^0$ and the $\omega$-type state with the $1\,^3D_1$
excitation, respectively. The predictions are consistent with some
experimental data but in strong contrast with the observations of
CMD-2 \cite{lite1} and CLEO \cite{lite13} which support the
$a_1(1260)$ dominance in the reaction
$e^+e^-\to\omega\pi$.

The prediction in the work \cite{lite12} that the meson
$\rho(1450)$ has a bigger probability to decay into $\omega\pi$
than the $\rho$-type mesons with higher masses is not consistent
with the results of the SND experiment \cite{lite8} that the
$\rho(1600)$ meson dominates over the $\rho(1400)$ in the reaction
$e^+e^-\to\omega\pi$. However, the results of the recent
work \cite{lite14} do not contradict the assignment of the
$\rho(1450)$ and $\omega(1420)$ to the state $2\,^3S_1$. We study
in the present work the reaction $e^+e^-\to\omega\pi$ at
low energies in the nonrelativistic $^3P_0$ quark model, aiming at
a better understanding of the reaction and the properties of the
vector mesons $\rho(1450)$ and $\rho(1700)$. The work is arranged as follows: In
Section II we study the reaction $e^+e^-\to\omega\pi$ in the
$^3P_0$ quark model and compare our results with experimental data.
Discussion and conclusions are given in Section III. The
transition amplitudes for the reaction of $\rho'$ to two S-wave mesons
are given in Appendix A while the model parameters are determined
in Appendix B and C.

\section{$e^+e^-\to\omega\pi$ in $^3P_0$ quark model}
The reactions $e^+e^-\to\omega\pi$ may stem from two
possible processes, namely, the one-step process where the
$e^+e^-$ pair annihilates into a virtual time-like photon, then the
virtual photon decays into a $\overline qq$ pair, and finally the
$\overline qq$ pair is dressed directly by an additional
quark-antiquark pair pumped out of the vacuum to form the
$\pi\omega$ final state, and the two-step process where the
$e^+e^-$ pair annihilates into a virtual time-like photon, then the
virtual photon decays into a $\overline qq$ pair, and the $\overline
qq$ pair first form a vector meson and finally the vector meson
decay into the $\omega\pi$ final state.

At high energies the reaction $e^+e^-\to\omega\pi$ is
likely dominated by the one-step process while in the low-energy
region, especially close to the threshold, the reactions are
expected to be dominated by the two-step process. It is found that
the reactions $e^+e^-\to\pi\pi, \overline NN$ at low
energies are dominated by the two-step process
\cite{lite16,lite17}. In this work we study the reaction
$e^+e^-\to\omega\pi$ at energies close to the threshold,
assuming that the reaction is dominated by the two-step
process.

The transition amplitude
of the reactions $e^+e^-\to\omega\pi$ in the two
step process shown in Fig. 1 takes the form
\begin{equation}\label{eq::9}
T=\langle\omega\pi| V_{\overline
qq}|\rho'\rangle\langle\rho'|G|\rho'\rangle\langle\rho'|\overline
qq\rangle\langle\overline qq|T|e^+e^-\rangle
\end{equation}
where $\langle\rho' |\overline qq\rangle$ is simply the wave
function of the intermediate meson $\rho'$, which takes the Gaussian
form in the work (see Appendix B). $\langle\rho'|G|\rho'\rangle$,
the Green function describes the propagation of the intermediate
meson, and $\langle\pi\omega|V_{\overline qq}|\rho'\rangle$ is the
transition amplitude of the intermediate meson $\rho'$ decaying to
the $\omega\pi$ pair in the $^3P_0$ nonrelativistic quark model.
Considering the energy region in question we include three
intermediate mesons $\rho(770)$, $\rho(1450)$ and $\rho(1700)$
\cite{lite3} in our calculation, and hence have the transition
amplitude take explicitly the form
\begin{eqnarray}\label{eq::9}
T & =&  \langle\omega\pi| V_{\overline
qq}|\rho(770)\rangle\langle\rho(770)|G|\rho(770)\rangle\langle\rho(770)|\overline
qq\rangle\langle\overline qq|T|e^+e^-\rangle \nonumber \\
& +& \langle\omega\pi| V_{\overline
qq}|\rho(1450)\rangle\langle\rho(1450)|G|\rho(1450)\rangle\langle\rho(1450)|\overline
qq\rangle\langle\overline qq|T|e^+e^-\rangle \nonumber \\
& +& \langle\omega\pi| V_{\overline
qq}|\rho(1700)\rangle\langle\rho(1700)|G|\rho(1700)\rangle\langle\rho(1700)|\overline
qq\rangle\langle\overline qq|T|e^+e^-\rangle
\end{eqnarray}

The transition amplitudes are evaluated, following the
conventions of the work \cite{lite17}, for the
process of the intermediate meson $\rho'$ decaying to two mesons. The explicit results are
given in Appendix A. The transition amplitude for the process $e^+e^-\to\rho'$
in eq. ({\ref{eq::9}) can be easily evaluated in the standard
method of quantum field theory, taking the form in eq.
(\ref{eq::5}). The Green function in Eq. (\ref{eq::9})
describing the propagation of the intermediate meson takes the
form
\begin{equation}
\langle\rho'|G|\rho'\rangle=\frac{1}{E_{cm}-(M_{\rho'}-i\,\Gamma_{\rho'}/2)}
\end{equation}
where $M_{\rho'}$ and $\Gamma_{\rho'}$ are the mass and width of
the intermediate meson $\rho'$, and $E_{cm}$ is the center-of-mass
energy of the system. The total cross section is finally obtained
from
\begin{equation}
    \sigma= \frac{2\pi E_1E_2\,k}{4\,E_{cm}^2\,p}\int|T_{e^+e^-\to m_1m_2}|^2 d\Omega
\end{equation}
where $E_i,\, E_{cm},\, p,\,k$ are respectively the energy of
the $i$th final meson, the center-of-mass energy of the system, the incoming
and outing momenta in the center of mass system.
\begin{figure}[h]
%\centerline{
%\epsfig{file=twostep.eps,width=3.in,height=2.in,angle=0}}
\centering
\includegraphics[width=0.3\textwidth]{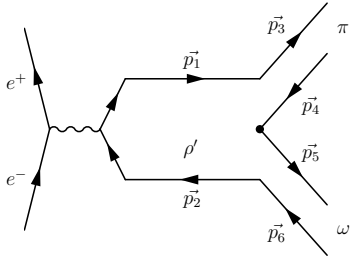}
\caption{\label{fig1} Reaction $e^+e^-\to\omega\pi$ in the two-step
process.}
\end{figure}
We intend to study the reaction $e^+e^-\to\omega\pi^0$ in the
$^3P_0$ nonperturbative quark model with all the model parameters
predetermined. There are two parameters for the $^3P_0$ quark model,
the meson size parameter and the effective coupling constant of the
$^3P_0$ vertex. In this work we take the meson size parameter
$a=3.24$ GeV$^{-1}$ as determined by the process
$\rho(\omega,\phi)\to e^-e^+$ in Appendix B, and the $^3P_0$
effective coupling constant $\lambda=1.25$ as determined by the
reaction $\rho\to\pi\pi$ in Appendix C. In the theoretical
calculation we have three intermediate mesons $\rho(770)$,
$\rho(1450)$ and $\rho(1700)$ included, with their masses and widths
taken from the particle data group \cite{lite3}. While the
$\rho(770)$ is kept always as a $1S$ meson, the $\rho(1450)$ and
$\rho(1700)$ are allowed to be the mixtures of the $2S$ and $1D$
states, that is
\begin{eqnarray}\label{mixture}
\rho(1450)=\cos\theta\,|2S\rangle+\sin\theta\,|1D\rangle \nonumber\\
\rho(1700)=-\sin\theta\,|2S\rangle+\cos\theta\,|1D\rangle
\end{eqnarray}
where $\theta$ is the mixing angle between the $2S$ and $1D$ states.

Shown in Fig. \ref{5.1} are the theoretical predictions and experimental data for the
cross section of the reaction $e^+e^-\to\omega\pi^0$. Note
that for comparing with the experimental data of the reaction
$e^+e^-\to\omega\pi^0\to\pi^0\pi^0\gamma$, we have
multiplied our theoretical predictions by the factor 0.087 which is
the decay branch ratio of $\omega(780)\to\pi^0\gamma$.
In the theoretical study the only free parameter is the mixing angle,
with the meson length parameter fixed to be 3.24 by the reaction $\rho\to e^+e^-$
and the effective coupling
constant of the $^3P_0$ quark vertex fixed to be 1.25 by the
reaction $\rho\to\pi\pi$. In Fig. \ref{5.1} the dotted curve represents the result with
$\rho(770)$ in the $1S$ state, $\rho(1450)$ in the $1D$ state and $\rho(1700)$ not included
while the dash-dotted curve is for the result with $\rho(770)$ in the $1S$ state, $\rho(1450)$
in the $1D$ state and $\rho(1700)$ in the $2S$ state. It is clear that the $\rho(1450)$ can not be
dominated by the $1D$ component and the $\rho(1700)$ is not allowed by the experimental data
to have a large $2S$ component.

The dashed curve is the theoretical result with $\rho(770)$ in the $1S$ state and $\rho(1450)$
and $\rho(1700)$ being respectively pure $2S$ and $1D$ mesons. The result is fairly consistent with
experimental data. Let $\rho(1450)$ and $\rho(1700)$ the mixtures of the $2S$ and $1D$ states, one may
improve the theoretical results to a certain extent. Shown in Fig. \ref{5.1} as the solid curve
is the theoretical result with $\rho(770)$ in the $1S$ state, $\rho(1450)$
and $\rho(1700)$ being the states in eq. (\ref{mixture}) with $\theta=\pi/6$. Considering the
uncertainty of the experimental data, however, it is difficult
to figure out how much D-wave component the $\rho(1450)$ may have.
\begin{figure}[h!]
%\begin{center}
  \centering
  \includegraphics[width=0.8\textwidth]{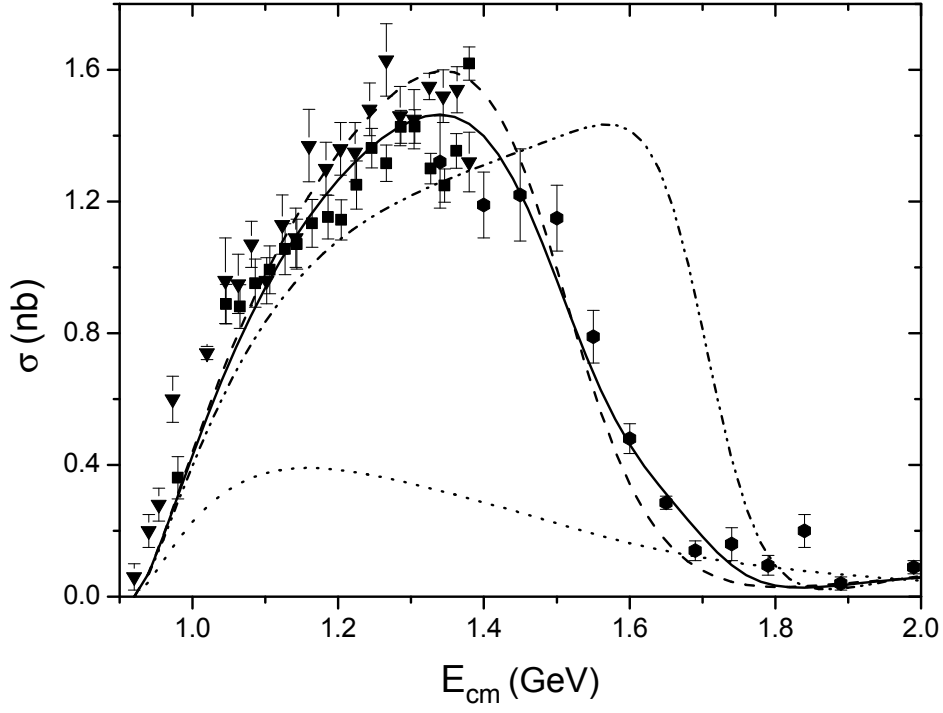}%\\
%  \vspace{-\baselineskip}
  \caption{\label{5.1} %
  Theoretical results for the cross section of the reaction $e^+e^-\to\omega\pi^0\to\pi^0\pi^0\gamma$:
  the dotted curve for $\rho(1450)$ in $2S$ state and $\rho(1700)$ not included, the dash-dotted
  curve for $\rho(1450)$ in $1D$ state and $\rho(1700)$ in $2S$ state,
  the dashed curve for $\rho(1450)$ in the $2S$ state and $\rho(1700)$ in $1D$ state,
  and the solid curve for the case where $\theta=\pi/6$ in eq. (\ref{mixture}).
  The experimental data are from the CMD2 \cite{lite1}, SND\cite{lite8} and DM2 \cite{lite19}}
%\end{center}
\end{figure}

\section{Discussion and Conclusions}
The reaction $e^+e^-\to\omega\pi^0$ in the energy region from the
$\omega\pi$ threshold to 2.0 GeV is
investigated in the $^3P_0$ nonrelativistic quark model with
model parameters predetermined by other processes.
The experimental data for the cross section of the reaction
$e^+e^-\to\omega\pi^0$ are well reproduced with $\rho(770)$, $\rho(1450)$
and $\rho(1700)$ as the intermediate states, where $\rho(770)$ is in the
$1S$ state, and $\rho(1450)$ and $\rho(1700)$ are dominantly in the
$2S$ and $1D$ states, respectively.

That the theoretical prediction for the cross section of the reaction
$e^+e^-\to\omega\pi^0$ in the two-step process with the
$\rho(770)$, $\rho(1450)$ and $\rho(1700)$ as the intermediate states
are consistent with the experimental data at energies below 2.0 GeV leaves no
room for the
one-step process to contribute to the reaction at a sizable scale
at this energy region. The study suggests
that at this energy region the two-step process
is dominant over the one-step one.

The work strongly suggest that the $\rho(1450)$ is
mainly a $2S$ meson and the $\rho(1700)$ a $1D$ one.

\section*{Acknowledgements}

The work is supported by the Commission on Higher Education, Thailand (CHE-RES-RG Theoretical Physics) and the corresponding author is supported in part by
the Development and Promotion of Science and Technology Talents Project (DPST), Thailand.

\appendix

\section{Transition of $\rho'\to\omega\pi$ in $^3P_0$ Model}
As the essential building blocks for the
transition amplitude of the reaction $e^+e^-\to\,\rho(\rho')\to\pi\omega$,
the transition amplitudes
for the processes $\rho'\to\omega\pi$ are derived in the $^3P_0$ quark model which are
described in details in Appendix C. For the purpose of good documentation, we
list obviously the amplitudes as follows:
\begin{align}\label{eq::16}
T_{\rho'\to\omega\pi} = f(k)\,Y_{1,j_z-m_\omega}(\hat k)
\end{align}
with
\begin{align}
    f(k)= A\,k\left(1-\frac{2a^2}{15}k^2\right)\,e^{-\frac{1}{12}a^2k^2}
\end{align}
where
\begin{align}
    A &= \lambda\,\frac{40\,\sqrt{2}\,a^{3/2}}{243\,\pi^{1/4}}{\mbox
    C}(111;j_z,m_{\omega}-j_z,m_{\omega})
\end{align}
for a $2S$ $\rho'$ decay to $\omega$ and $\pi$ mesons, and
\begin{align}
    A &=
    \lambda\,\frac{16\,\sqrt{10}\,a^{3/2}}{243\,\pi^{1/4}}
    {\mbox C}(111;m_{\rho'},m_\omega-m_{\rho'},m_\omega){\mbox C}(121;m_{\rho'},j_z-m_{\rho'},j_z)
    {\mbox C}(121;m_\omega-j_z,j_z-m_{\rho'},m_\omega-m_{\rho'})
\end{align}
for a $1D$ $\rho'$ decay to $\omega$ and $\pi$ mesons. Here in the
above equations $m_{\rho'}$ and $m_\omega$ are respectively the spin
magnetic quantum numbers of the intermediate $\rho'$ meson and the
final $\omega$ meson, and $j_z$ the total magnetic quantum number of
the system. The C-G coefficients above are in the form
$C(l_1l_2l_3;m_1m_2m_3)$. In the evaluation the meson spatial wave
functions have been taken as the Gaussian form in eq.(\ref{swave}).
The transition amplitudes above differ by a global factor from
the ones in the work \cite{barnes} since we have defined different color matrices
for the decay processes in question.
\section{Meson size parameter}
Our intention is to study the reaction $e^+e^-\to\omega\pi^0$ in
the $^3P_0$ nonperturbative quark
model with all the model parameters predetermined. In this study there
are two model parameters, the meson size parameter and the effective
coupling constant of the $^3P_0$ vertex. The meson size parameter
comes with the meson spatial wave functions, which take
the Gaussian form as usual,
\begin{equation}
    R_{nl}(p)=\left[\frac{2a^3n!}{\Gamma(n+l+\frac{3}{2})}\right](ap)^l
    e^{-\frac{1}{2}a^2r^2}L_n^{l+1/2}(a^2p^2).
\end{equation}
where $L_n^{l+1/2}(\alpha^2r^2)$ are the associated Laguerre
polynomials, $p$ is the magnitude of the relative momentum between the quark
and antiquark, and $n$, $l$ and $a$ are respectively the principle quantum number,
orbital quantum number and the mentioned size parameter. For the lowest states
$1S$, $2S$ and $1D$, we have the meson spatial wave functions
\begin{equation}\label{swave}
    \psi_{sp}=\frac{\sqrt{a^3}}{\pi^{3/4}}e^{-\frac{1}{2}a^2p^2},
\end{equation}
\begin{equation}
    \psi_{sp}=\frac{1}{\pi^{3/4}}\sqrt{\frac{2a^3}{3}}\left(\frac{3}{2}-a^2p^2\right)e^{-\frac{1}{2}a^2p^2},
\end{equation}
and
\begin{equation}
    \psi_{sp}=\frac{4}{\pi^{1/4}}\sqrt{\frac{a^3}{15}}~(ap)^2e^{-\frac{1}{2}a^2p^2}Y_{2m}(\theta,\phi)
\end{equation}

The size parameter may be determined by the
reaction $\rho\,(\omega,\phi)\to e^+e^-$. The transition amplitude of
the reaction is
\begin{align}\label{eq::5}
    T_{\rho\to e^+e^-}&=\sum_\alpha\sum_{s_qs_{\bar q}}\sum_{t_qt_{\bar
    q}}\frac{1}{\sqrt3}~C(\frac{1}{2}\frac{1}{2}S,s_qs_{\bar
    q}S_z)C(\frac{1}{2}\frac{1}{2}I,t_qt_{\bar
    q}I_z)\nonumber\\
    &\quad\cdot\int\frac{d\vec
    p}{(2\pi)^{3/2}2E_q}~\psi_\rho(\vec p)~T_{q\bar q\to e^+e^-}
\end{align}
where $\alpha$ is color indices of the quark and antiquark of
the $\rho$ meson, $s_q (t_q)$ and $s_{\bar q} (t_{\bar q})$ are the
z-axis spin (isospin) projections for quark and antiquark, respectively.
$S (I)$ (actually equal to 1) and $S_z (I_z)$ are respectively the spin (isospin)
and the z-axis
spin (isospin) projection of the $\rho$ meson. $\psi_\rho(\vec
p)$ is the spatial wave function of the $\rho$ meson in momentum space with
$\vec p\,$ being the relative momentum between the quark and
antiquark inside, and $T_{q\bar q\to e^+e^-}\equiv\langle
e^+e^-|T|q\bar q\rangle$ is the transition amplitude of the reaction
of a quark-antiquark pair to an electron-positron pair, taking the
form
\begin{equation}\label{eq::6}
\langle e^+e^-|T|q\bar q\rangle=-\frac{e_qe}{s}\bar
    u_e(p_{e^-},m_{e^-})\gamma^\mu v_e(p_{e^+},m_{e^+})
    \bar v_q(p_{\bar q},m_{\bar q})\gamma_\mu
    u_q(p_{q},m_{q})
\end{equation}
where $s=(p_q+p_{\bar q})^2$, $e_q$ is the quark charge, and the
Dirac spinors are normalized according to $\bar uu=\bar vv=2m_q$.

We derive the size parameter $a=3.32$, 3.49 and 2.92 GeV$^{-1}$ from the reactions $\rho\to e^+e^-$,
$\omega\to e^+e^-$ and $\phi\to e^+e^-$, respectively. Here we have used as inputs
$\alpha=1/137$, $M_\rho=0.775$ GeV, $M_\omega=0.782$ GeV, $M_\phi=1.019$ GeV,
the quark masses $m_u=m_d=5$ MeV and $m_s=100$ MeV,
and the experimental
values of $\Gamma_{\rho^0\to e^+e^-}=7.02$ keV,
$\Gamma_{\omega\to e^+e^-}=0.60$ keV and $\Gamma_{\phi\to e^+e^-}=1.27$ keV.
The averaged value of the meson size parameter is $a=3.24$ GeV$^{-1}$.

The transition amplitude of the decay process $\rho\to e^+e^-$ in eq. (\ref{eq::5}) is
made up of two pieces: the transition amplitude $T_{q\bar q\to e^+e^-}$ of the process $q\bar q \to e^+e^-$ as shown
in eq. (\ref{eq::6}), and the projection of the quark-antiquark pair to the $\rho$ meson state,
which is performed by the summation and integration in eq. (\ref{eq::5}).
The masses of quark and antiquark appear only in the transition amplitude
$T_{q\bar q\to e^+e^-}$ and the spatial
wave function $\psi_\rho(\vec p)$ of the $\rho$ meson in momentum space has no
direct dependence of quark masses. Therefore, it may be more reasonable to let quarks
take the current masses instead of the constituent masses.

\section{Strength of $^3P_0$ quark model}
In this work we study the reaction $e^+e^-\to\omega\pi^0$ in the nonperturbative quark
model with the $^3P_0$ quark dynamics which describes the
quark-antiquark annihilation and creation. The $^3P_0$ model,
first introduced in the work \cite{lite15}, has made considerable
successes in understanding low-energies hadron physics \cite{3P0,tueb1,dgmf,muhn}.

The $^3P_0$ decay model defines the
quantum states of a quark-antiquark pair destroyed into or
created from vacuum to be $J=0, L=1, S=1$ and $T=0$.
The effective vertex in $^3P_0$ model takes the form
\begin{align} \label{eq::1}
    V_{ij}&=\lambda\vec\sigma_{ij}\cdot(\vec p_i-\vec
    p_j)\hat{F}_{ij}\hat{C}_{ij}\delta(\vec p_i+\vec
    p_j)\nonumber\\
    &=\lambda\sum_\mu\sqrt\frac{4\pi}{3}(-1)^\mu\sigma_{ij}^\mu Y_{1\mu}
    (\vec p_i-\vec p_j)\hat F_{ij}\hat C_{ij}\delta(\vec p_i+\vec
    p_j)
\end{align}
where $\sigma_{ij}^\mu, \hat F_{ij}, \hat C_{ij},$ and $\lambda$
are respectively the spin, flavor and color operators,
and the effective coupling constant. The operations of flavor, color, and
spin operators onto a $q\overline q$ pair are
\begin{align}\label{eq::2}
\langle 0,0|\hat{F}_{ij}|T,T_z\rangle &= \sqrt
    2\delta_{T,0}\delta_{T_z,0}, \nonumber \\
\langle 0,0|\hat{C}_{ij}|q_\alpha^i\bar q_\beta^j\rangle &=
    \delta_{\alpha\beta}, \nonumber \\
\langle
0,0|\sigma_{ij}^\mu|\left[\overline\chi_i\otimes\chi_j\right]_{JM}\rangle&=
    (-1)^M\sqrt 2\delta_{J,1}\delta_{M,-\mu}
\end{align}
where $\chi_i (\overline\chi_i)$ is the spin state of quark (antiquark), $\alpha$ and $\beta$ the color indices,
and $T$ and $T_z$ the total isospin and its projection of the $q\overline q$.

In this work we determine the effective coupling constant of the $^3P_0$ quark vertex with the reaction
$\rho(770)\to\pi^+\pi^-$.
The decay width of the reaction $\rho\to\pi^+\pi^-$ takes the form
\begin{align}\label{eq::3}
    \Gamma_{\rho^0\to
    \pi^+\pi^-}=\frac{\pi}{4}M^2_\rho\sqrt{1-\frac{4M^2_\pi}{M^2_\rho}}|T_{\rho^0\to
    \pi^+\pi^-}|^2
\end{align}
In the $^3P_0$ model, the transition amplitude
$T_{\rho^0\to\pi^+\pi^-}$ is derived as in the
center-of-mass system \cite{lite17},
\begin{equation}\label{eq::4}
T_{\rho^0\to\pi^+\pi^-}=-
    \frac{16\lambda~a^{3/2}e^{-\frac{1}{12}a^2k^2}k}{27\sqrt3~\pi^{1/4}}
\end{equation}
where $a$ is the size parameter of the mesons, which appears in
the wave functions of the mesons. $k$ in eq. (\ref{eq::4}) is the
momentum of the final $\pi$ mesons. The transition amplitude differ by a constant
factor from the one in the work \cite{barnes} since we have used different color matrix.
Using as an input $a=3.24$ GeV$^{-1}$ and experimental data for $M_\pi$, $M_\rho$ and
$\Gamma_{\rho^0\to\pi^+\pi^-}$ taken from \cite{lite3} to eq. (\ref{eq::4}), one derives
$\lambda=1.25$.

 \end{document}